\documentclass[preprint,preprintnumbers,amsmath,amssymb,color]{revtex4}
\begin{document}

\title{Violation of the first law of black hole thermodynamics in $f(T)$ gravity}

\author{Rong-Xin Miao$^{a}$}
\author{Miao Li$^{b}$}
\author{Yan-Gang Miao$^{c}$}

\affiliation{$^{a}$Interdisciplinary Center for Theoretical Study, University of Science and Technology of China,\\
Hefei, Anhui 230026, People's Republic of
China.}\email{mrx11@mail.ustc.edu.cn}

\affiliation{$^{b}$Kavli Institute for Theoretical Physics, Key
Laboratory of Frontiers in Theoretical Physics, Institute of
Theoretical Physics, Chinese Academy of Sciences, \\
Beijing 100190, People's Republic of China. }\email{mli@itp.ac.cn}

\affiliation{$^{c}$School of Physics, Nankai University, Tianjin
300071, People's Republic of China.}\email{miaoyg@nankai.edu.cn}

\preprint{USTC-ICTS-11-08}

\begin{abstract}
We prove that, in general, the first law of black hole
thermodynamics, $\delta Q=T\delta S$, is violated in $f(T)$ gravity.
As a result, it is possible that there exists entropy production,
which implies that the black hole thermodynamics can be in
non-equilibrium even in the static spacetime. This feature is very
different from that of $f(R)$ or that of other higher derivative
gravity theories. We find that the violation of first law results
from the lack of local Lorentz invariance in $f(T)$ gravity. By
investigating two examples, we note that $f''(0)$ should be negative
in order to avoid the naked singularities and superluminal motion of
light. When $f''(T)$ is small, the entropy of black holes in $f(T)$
gravity is approximatively equal to $\frac{f'(T)}{4}A$.
\end{abstract}

%\pacs{11.30.Cp, 04.50.Kd, 04.80.Cc, 11.30.Qc}% PACS, the Physics and Astronomy
                             % Classification Scheme.
\maketitle

\section{Introduction}
 $f(T)$ gravity as a new modified gravity theory has recently
attracted much attention
 \cite{Ferraro1,Ferraro2,Bengochea:2009,Linder:2010, Yu:2010a, Myrzakulov:2010a, Tsyba:2010,
  Yi:2010b, Kazuharu:2010a, Kazuharu:2010b, Myrzakulov:2010b,
 Yu:2010c, Karami:2010, Dent:2010, Dent:2011,Wei,
 Sotiriou:2010, Li, Li2, Tower,Daouda1,Daouda2, Cai,Ferraro3,Ferraro4,Deliduman,Miao}.
It was first investigated by Ferraro and Fiorini
\cite{Ferraro1,Ferraro2} in the Born-Infeld style which can lead to
regular cosmological spacetimes without Big Bang singularity. Then,
it was proposed by Bengochea, Ferraro and Linder
\cite{Bengochea:2009,Linder:2010} to explain the current accelerated
expansion of universe. Similar to $f(R)$ gravity,
 it is a generalization of the teleparallel gravity ($TG$) \cite{ein28,eineng,Aldrovandi} which was originally developed
 by Einstein in an attempt of unifying gravity and
 electromagnetism. Let us make a brief review of $TG$. The basic variables in $TG$ are tetrad fields
 $e_{a\mu}$, where $a$ is index of the internal space running over
 $0, 1, 2, 3$ while $\mu$ is the spacetime index running from
 0 to 3. The tetrad fields are related with the spacetime metric by
\begin{eqnarray}\label{tetrad}
g_{\mu\nu}=e_{a\mu}\eta^{ab}e_{b\nu}, \qquad
\eta_{ab}=e_{a\mu}e_{b\nu}g^{\mu\nu}={\rm diag}(-1,1,1,1).
\end{eqnarray}
In $TG$, the Weitzenbock connection
\begin{eqnarray}\label{Weizenbock}
\Gamma^{\lambda}_{\ \mu\nu}=e_a^{\lambda}\partial_\nu e^a_\mu
\end{eqnarray}
rather than the Levi-Civita connection is used to define the
covariant derivative, and as a result there is no curvature but only
torsion
\begin{eqnarray}\label{torsion}
T^{\lambda}_{\ \mu\nu}=\Gamma^{\lambda}_{\
\nu\mu}-\Gamma^{\lambda}_{\ \mu\nu}=e^{a\lambda}(\partial_\mu
e_{a\nu}-\partial_\nu e_{a\mu}).
\end{eqnarray}
The torsion scalar is defined by
\begin{eqnarray}\label{torsionscalar}
T=\frac{1}{2}S^{\mu\nu\rho}T_{\mu\nu\rho}=\frac{1}{4}T^{\mu\nu\rho}T_{\mu\nu\rho}+\frac{1}{2}T^{\mu\nu\rho}T_{\rho\nu\mu}-T_{\sigma}^{\
\sigma\mu}T^{\rho}_{\rho\mu},
\end{eqnarray}
with the so-called dual torsion
\begin{eqnarray}\label{dualtorsion}
S^{\mu\nu\rho}=\frac{1}{2}(T^{\mu\nu\rho}+T^{\nu\mu\rho}-T^{\rho\mu\nu})+g^{\mu\rho}T_{\sigma}^{\
\sigma\nu}-g^{\nu\rho}T_{\sigma}^{\ \sigma\mu}.
\end{eqnarray}
There are several virtues in $TG$. For example, in contrast to
Einstein gravity, a covariant stress tensor of gravitation can
naturally be
 defined in the gauge context of $TG$ \cite{Andrade}.

 As a main advantage compared with $f(R)$
 gravity, the equations of motion of $f(T)$ gravity are second-order instead of
 fourth-order. However, the local Lorentz invariance is violated in $f(T)$ gravity
\cite{Li} and consequently more degrees of
 freedom appear. Recently, we have investigated the Hamiltonian formulation of $f(T)$ gravity and have
found that three extra degrees of freedom emerge \cite{Miao}. In
general, there are $D-1$ extra degrees of freedom for $f(T)$ gravity
in $D$ dimensions, and this implies that the extra degrees of
freedom might correspond to one massive vector field. For the
detailed explanation, see our recent work~\cite{Miao}.

In this paper, we investigate the black hole thermodynamics in
$f(T)$ gravity and find that the first law, $\delta Q=T\delta S$, is
violated. There is entropy production even in the static spacetime
and the black hole thermodynamics turns out to be non-equilibrium.
By analyzing two examples in detail, we find that it is the
violation of the local Lorentz invariance in $f(T)$ gravity that
leads to the breakdown of the first law of black holes. Because of
this violation, some degrees of freedom in $f(T)$ gravity feel an
effective metric different from the background metric. Consequently,
they see a different horizon and Hawking temperature from that felt
by matter fields with the local Lorentz invariance. Black holes in
such a situation would not be in equilibrium, thus it is not
surprising that the first law is violated. In addition, from the two
examples that will be analyzed, we also observe that $f''(0)$ should
be negative in order to avoid the naked singularities and super
velocity of light.

It should be stressed that, by ``black hole'' in $f(T)$ gravity, we
mean in the sense of the usual metric. Recently, some ``black
holes'' in this sense were found in \cite{Ferraro5}. In general,
there may exist modes which can escape from the inside and make the
horizon defined by the metric ``non-black''. However, as shown in
Appendix B, there indeed exist exact solutions of $f(T)$ gravity
which have the properties of the usual black hole. All the modes
feel the same metric and no modes can escape from the inside of the
horizon. We focus on the ``black hole'' in the metric in this paper.

The paper is arranged as follows. In Sect. II, we give a brief
 review of the first law of $f(R)$ gravity using the field equation method. In Sect. III, we establish the first law of
 $f(T)$ gravity. In Sect. IV, we search for the reasons for the
 violation of first law of $f(T)$ gravity by investigating two examples.
 We conclude in Sect. V.

\section{First law of $f(R)$ gravity}

The first law of black holes, $\delta Q=T\delta S$, is universal for
gravity with the diffeomorphism Lagrangian, $L(g_{\mu\nu},
R_{\mu\nu\rho\sigma})$, constructed from the metric $g_{\mu\nu}$ and
Riemann tensor $R_{\mu\nu\rho\sigma}$. One can derive the first law
and entropy of black holes from various procedures, for instance,
the Wald's Noether charge method \cite{Wald}. However, we shall use
a different approach \cite{Jacobson,Jacobson1,Brustein,MiaoLi} which
was originally developed to derive the gravity field equations from
the thermodynamic point of view. In this paper we turn the logic
around: we suppose the gravity field equations and check if the
thermodynamic relation $\delta Q=T\delta S$ is satisfied. Though
similar in some aspects, there are many differences between the
Wald's Noether charge approach and the field equation approach. Here
we just list three main differences. First, the Wald's Noether
charge approach is based on the Lagrangian or the action of a
theory, while the field equation approach is based on the equations
of motion. Second, the definitions of energy are different in the
two approaches. In the former, Wald uses the ``canonical energy''
$E$ from which one can derive $\delta E=T\delta S+\Omega_H \delta
J$, where $\Omega_H$ is the angular velocity of the horizon and $J$
is the angular momentum. In the latter, one defines the heat flux
passing through the null surface as eq.~(\ref{Q1}), which does not
contain the information of angular momentum. As we shall show below,
using eq.~(\ref{Q1}), one can only derive $\delta Q=T\delta S$.
Third, it is natural to use the field equation approach rather than
the Wald's approach to study the first law of black holes in $f(T)$
gravity. The Wald's approach is not designed for the teleparallel
gravity. The key point of the field equation approach is the
definition of the heat flux passing through the null surface.
According to ref.~\cite{Li}, we still have $\nabla^\mu T_{\mu\nu}=0$
in $f(T)$ gravity, and therefore the current $T_{\mu\nu}\xi^{\mu}$
remains conserved, i.e. $\nabla^\mu(T_{\mu\nu}\xi^{\nu})=0$. Thus,
it is very natural to use eq.~(\ref{Q1}) as the heat flux passing
through the null surface in $f(T)$ gravity.

Now we give a brief review of the field equation approach. Let us
take $f(R)$ gravity as an example, and consider a heat flux $\delta
Q$ passing through an open patch on a null surface or black hole
horizon, $dH=dAd\lambda$,
\begin{equation}\label{Q1}
 \delta Q=\int_{H} T_{\mu\nu}\xi^{\mu}k^{\nu}dA d\lambda ,
\end{equation}
where $T_{\mu\nu}$ is the matter stress-tensor, $\xi^\mu$ is the
Killing vector, $H$ denotes the null surface, $\lambda$ is the
affine parameter, and $k^\mu=\frac{dx^\mu}{d\lambda}$ is the tangent
vector to $H$.
 Substituting the $f(R)$ field
equation
\begin{eqnarray}\label{f(R)}
f'(R)R_{\mu\nu}-\nabla_\mu\nabla_\nu f'(R)+g_{\mu\nu}\biggl(\Box
f'(R)-\frac{1}{2}f(R)\biggr)=8\pi T_{\mu\nu}
\end{eqnarray}
into eq.~(\ref{Q1}), we can derive
\begin{eqnarray}\label{Q2}
\delta Q&=&\frac{1}{8\pi}\int_{H}
\biggl(f'(R)R_{\mu\nu}-\nabla_\mu\nabla_\nu
f'(R)\biggr)\xi^{\mu}k^{\nu}dA d\lambda\nonumber\\
&=&\frac{1}{8\pi}\int_{H}
\biggl(f'(R)\nabla_\mu\nabla_\nu\xi^\mu-\xi^\mu\nabla_\mu\nabla_\nu
f'(R)\biggr)k^{\nu}dA d\lambda\nonumber\\
&=&\frac{1}{8\pi}\int_{H}
\biggl(k^{\nu}\nabla^\mu\big(f'(R)\nabla_\nu\xi_\mu\big)\biggr)dA d\lambda\nonumber\\
&=&\frac{1}{8\pi}\int_{H}
\biggl(k^{\nu}l^\mu f'(R)\nabla_\nu\xi_\mu\biggr)dA d\lambda\nonumber\\
&=&\frac{\kappa}{2\pi}\Big(\frac{f'(R)dA}{4}\Big)\Big|^{d
\lambda}_0=T\delta S.\label{firstlaw}
\end{eqnarray}
In the above derivations, we have used Stokes's Theorem and the
following formulas:
\begin{eqnarray}\label{formula1}
k^\mu\xi_\nu=0,\qquad  k^\mu k_\mu=0,\qquad  l^\mu l_\mu=0, \qquad
k^\mu l_\mu=-1,
\end{eqnarray}
\begin{eqnarray}\label{formula2}
R_{\mu\nu}\xi^\mu=\nabla_\mu\nabla_\nu\xi^\mu,\qquad
\xi^\mu\nabla_\mu R=0,
\end{eqnarray}
\begin{eqnarray}\label{formula3}
k^\mu l^\nu \nabla_\mu \xi_\nu=\kappa,\qquad
T=\frac{\kappa}{2\pi},\qquad \frac{d \kappa}{d\lambda}=0,
\end{eqnarray}
where $\kappa$ is the surface gravity of the null surface $H$. From
eq.~(\ref{Q2}), we can read out the entropy of black holes as
$S=\frac{f'(R)A}{4}$, which is consistent with the Wald's result.

It should be stressed that in order to derive the first law, $\delta
Q=T\delta S$, in eq.~(\ref{Q2}), we have used the formula
eq.~(\ref{formula2}) which is valid only for an exact Killing vector
$\xi$. However, in general, there is no such an exact Killing vector
in a dynamic spacetime. One can at most obtain a Killing vector to
the second order (in Riemann normal coordinates), $\xi^\mu=-\lambda
k^\mu + \mathcal{O}(\lambda^3)$, in our case \cite{Jacobson}. Lack
of an exact Killing vector implies that the spacetime might be out
of equilibrium and leads to the appearance of extra terms in
eq.~(\ref{Q2}), which can be explained as contributions from entropy
production in view of Jacobson's idea \cite{Jacobson1}.

For simplicity, we focus on the cases with exact Killing vectors
below. Note that the static and stationary black holes always have
an exact Killing vector, therefore our discussions are universal
enough. The method mentioned above can easily be generalized to
gravity with the diffeomorphism Lagrangian, $L(g_{\mu\nu},
R_{\mu\nu\rho\sigma})$, constructed from the metric and Riemann
tensor. Substituting the field equation
\begin{eqnarray}\label{Ggravity}
P_a ^{\ cde}R_{bcde}-2\nabla^c\nabla^d
P_{acdb}-\frac{1}{2}Lg_{ab}=8\pi T_{ab},\qquad
P^{abcd}=\frac{\partial L}{\partial R_{abcd}}
\end{eqnarray}
into eq.~(\ref{Q1}), one can derive
\begin{eqnarray}\label{Q3}
\delta Q=\frac{1}{8\pi}\Big(k_al_b
\left(P^{abcd}\nabla_c\xi_d-2\xi_d\nabla_cP^{abcd}\right)dA\Big)\Big|^{d
\lambda}_0\Big.=T\delta S,
\end{eqnarray}
where
$S=\frac{1}{4\kappa}(P^{abcd}\nabla_c\xi_d-2\xi_d\nabla_cP^{abcd})k_al_bdA$
is equivalent to Wald entropy \cite{Parikh}.

\section{Violation of First law of $f(T)$ gravity}

Now we use the field equation method introduced in Sec. II to
investigate the first law of black hole thermodynamics in $f(T)$
gravity. We find that the Clausius relation, $dS=\frac{dQ}{T}$, is
violated, which implies that even in a static spacetime the black
hole of $f(T)$ gravity is out of equilibrium and gives an intrinsic
entropy production.

Let us recall the equation of motion of $f(T)$ gravity \cite{Li},
\begin{eqnarray}\label{motion1}
H_{\mu\nu}=f'(T)(R_{\mu\nu}-\frac{R}{2}
g_{\mu\nu})+\frac{1}{2}g_{\mu\nu}[f(T)-f'(T)T]+f''(T)S_{\nu\mu\rho}\nabla^\rho
T=8\pi\Theta_{\mu\nu},\label{eom}
\end{eqnarray}
\begin{eqnarray}\label{motion10}
H_{[\mu\nu]}=f''(T)S_{[\nu\mu]\rho}\nabla^\rho T=0,
\end{eqnarray}
where $\Theta_{\mu\nu}$ is the matter stress-tensor. As the matter
action is supposed to be invariant under the local Lorentz
transformation, $\Theta_{\mu\nu}$ is symmetric and satisfies
$\nabla^\mu\Theta_{\mu\nu}=0$. Notice that eq.~(\ref{motion10}) is
just the antisymmetric part of eq.~({\ref{motion1}). According to
ref.~\cite{Li}, eqs.~({\ref{motion1}) and (\ref{motion10}) are not
Lorentz invariant. This leads to an important fact that the solution
of  eqs.~({\ref{motion1}) and (\ref{motion10}) is unique for every
given $\Theta_{\mu\nu}$. Unlike Einstein gravity or $T$ gravity, in
general, one cannot get a new solution of  eqs.~({\ref{motion1}) and
(\ref{motion10}) from the old one by performing local Lorentz
transformations.

The Hawking radiation is known to be independent of dynamics of
gravity, which is a purely kinematic effect that is universal for
Lorentz geometries containing an event horizon \cite{Visser}. Thus,
the Hawking temperature felt by matter (whose action has a local
Lorentz invariance) in $f(T)$ gravity is the same as that in
Einstein gravity, $T=\frac{\kappa}{2\pi}$. On the other hand, the
entropy of black holes is related to dynamics of gravity. Now let us
begin to study the first law and entropy of black holes in $f(T)$
gravity, we still focus on the spacetime with an exact Killing
vector. By ``Killing vector $\xi^\mu$'', we mean in the sense of the
usual metric that it satisfies the equation, $\mathcal{L}_\xi
g_{\mu\nu}=\xi^{\alpha}\partial_\alpha
g_{\mu\nu}+\partial_\mu\xi^{\alpha}g_{\alpha\nu}+\partial_\nu\xi^{\alpha}g_{\alpha\mu}=0$.
Since one metric corresponds to many different tetrad fields which
are related with each other by local Lorentz transformations, it is
possible that the metric is static while the tetrad fields are time
dependent

Consider a heat flux $\delta Q$ passing through an open patch on a
null surface or black hole horizon, we have
\begin{equation}\label{Q4}
 \delta Q=\int_{H} \Theta_{\mu\nu}\xi^{\mu}k^{\nu}dA d\lambda.
\end{equation}
Substituting eq.~(\ref{eom}) into the above equation, we get
\begin{eqnarray}\label{Q5}
 \delta Q&=& \frac{1}{8\pi}\int_{H} k^{\nu}[f'(T)R_{\mu\nu}\xi^{\mu}+\xi^{\mu}S_{\nu\mu\rho}\nabla^\rho f'(T)]dAd\lambda\nonumber \\
 &=& \frac{1}{8\pi}\int_{H} k^{\nu}[f'(T)\nabla_{\mu}\nabla_{\nu}\xi^{\mu}+\xi^{\mu}S_{\nu\mu\rho}\nabla^\rho f'(T)]dAd\lambda\nonumber \\
 &=& \frac{1}{8\pi}\int_{H} k^{\nu}[\nabla^{\mu}(f'(T)\nabla_{\nu}\xi_{\mu})
 -(\nabla^{\mu}f'(T))\nabla_{\nu}\xi_{\mu}+\xi^{\mu}S_{\nu\mu\rho}\nabla^\rho f'(T)]dAd\lambda\nonumber \\
 &=&\frac{\kappa}{2\pi}\Big(\frac{f'(T)dA}{4}\Big)\Big|^{d\lambda}_0
 +\frac{1}{8\pi}\int_{H} k^{\nu}\nabla^\mu f'(T)(\xi^\rho S_{\rho \nu\mu}-\nabla_\nu \xi_\mu)dA
 d\lambda.
\end{eqnarray}
Note that in the above derivations, we have used
$R_{\mu\nu}\xi^\mu=\nabla_\mu\nabla_\nu\xi^\mu$ and $\xi^\mu\sim
k^\mu$ on the null surface, and thus we have $\xi^\mu k^\nu
S_{\mu\nu\rho}=\xi^\nu k^\mu S_{\mu\nu\rho}$. It should be mentioned
that since $\xi^\mu\sim k^\mu$ on the null surface, only the
symmetrical part of eq.~(\ref{motion1}) contributes to
eq.~(\ref{Q5}), while the antisymmetric part eq.~(\ref{motion10})
does not contribute to eq.~(\ref{Q5}).

The first term
$\frac{\kappa}{2\pi}(\frac{f'(T)dA}{4})|^{d\lambda}_0$ in the above
equation is similar to the last line of eq.~(\ref{firstlaw}),
therefore it can be explained as $T\delta S$. It is interesting that
an extra term appears which in general neither vanishes nor can be
rewritten in the form $\int_{H} k^{\nu}\nabla^\mu
B_{[\nu\mu]}dAd\lambda$ for an arbitrary $f'(T)$. We shall give the
proof below.

If the second term vanishes for an arbitrary  $f'(T)$, we then get
$k^{\nu}\xi^\rho S_{\rho \nu\mu}-k^{\nu}\nabla_\nu \xi_\mu=0$.
However, due to the fact that $k^{\nu}\nabla_\nu \xi_\mu$ is a
Lorentz scalar but $k^{\nu}\xi^\rho S_{\rho \nu\mu}$ is not, the two
terms can not be equal to each other. This contradiction shows that
the second term of eq.~(\ref{Q5}) is non-vanishing. Similarly,
suppose that the second term can be rewritten as $k^{\nu}\nabla^\mu
B_{[\nu\mu]}$ for an arbitrary $k^{\mu}$ (we can change the
direction of $k^{\mu}$ arbitrarily by choosing a different open
patch of the null surface or choosing a different null surface), we
have $\nabla^\mu B_{[\nu\mu]}=\nabla^\mu f'(T)(\xi^\rho S_{\rho
\nu\mu}-\nabla_\nu \xi_\mu)$. Note that $\nabla^\nu\nabla^\mu
B_{[\nu\mu]}=R^{\mu\nu}B_{[\nu\mu]}=0$, we can obtain $\nabla^\mu
f'(T)\nabla^\nu(\xi^\rho S_{\rho \nu\mu}-\nabla_\nu \xi_\mu)=0$. For
an arbitrary $f'(T)$ we deduce $\nabla^\nu(\xi^\rho S_{\rho
\nu\mu}-\nabla_\nu \xi_\mu)=0$. Considering the fact that
$\nabla^\nu\nabla_\nu \xi_\mu$ is a local Lorentz scalar while
$\nabla^\nu(\xi^\rho S_{\rho \nu\mu})$ is not, we conclude that the
second term of eq.~(\ref{Q5}) would not take the form
$k^{\nu}\nabla^\mu B_{[\nu\mu]}$. Notice that we do not use
eq.~(\ref{motion10}) in the above derivations. One may guess that
the last term of eq.~(\ref{Q5}) vanishes provided
eq.~(\ref{motion10}) is used. However, it is not the case. There
exist tetrad fields that satisfy eqs.~(\ref{motion1}) and
(\ref{motion10}) but still make the last line of eq.~(\ref{Q5})
non-vanishing. To end up the proof, we give an example in Appendix A
to show that the second term in the last line of eq.~(\ref{Q5}) is
indeed non-vanishing even provided eq.~(\ref{motion10}) is used.

It should be mentioned that we have proved that, in general, the
first law of black bole thermodynamics is violated for $f(T)$
gravity. But there might exist some special cases in which the first
law of $f(T)$ black boles recovers. Note that for black holes with
the same metric $g_{\mu\nu}$, we have many different choices of
tetrad fields $e_{a\mu}$ which are related with each other by local
Lorentz transformations. Those black holes have the same
$k^{\nu}\nabla_\nu \xi_\mu$ but different $k^{\nu}\xi^\rho S_{\rho
\nu\mu}$. Thus, for some special cases, the two terms might cancel
each other and the second term of the last line of eq.~(\ref{Q5})
vanishes. We give such an example in Appendix B in which the first
law $\delta Q=T\delta S$ recovers on the null surface.

Similar to $f(R)$ gravity \cite{Jacobson1}, the second term of
eq.~(\ref{Q5}) may be explained as contributions from entropy
production
\begin{eqnarray}\label{entropy pro}
\frac{1}{8\pi}\int_{H} k^{\nu}\nabla^\mu f'(T)(\xi^\rho S_{\rho
\nu\mu}-\nabla_\nu \xi_\mu)dA
 d\lambda =-T\delta S_i,
\end{eqnarray}
which implies the black hole thermodynamics becomes non-equilibrium,
$\delta Q = T\delta S-T\delta S_i$.

It should be stressed that there is one main difference between the
entropy production of $f(R)$ gravity and that of $f(T)$ gravity. For
$f(R)$ gravity, when the Killing vector is exact (for example, the
static and stationary black holes), the entropy production vanishes.
While for $f(T)$ gravity, we find that there is entropy production
even in a static spacetime. Since the entropy and entropy production
should always be positive, there are very strict constraints for
$f(T)$ gravity,
\begin{eqnarray}\label{constraint}
f'(T)>0,\qquad f''(T)k^{\nu}\nabla^\mu T(\xi^\rho S_{\rho
\nu\mu}-\nabla_\nu \xi_\mu) \leq 0.
\end{eqnarray}

The local Lorentz invariance has been examined by experiment in many
sectors of the standard model, including photons, electrons, protons
and neutrons \cite{Lorentz1,Lorentz2,Lorentz3}. No violation of
Lorentz symmetry has been identified so far in these sectors.
M\"{u}ller et al. performed an experiment to test the local Lorentz
symmetry in the gravitational sector and they found a small
violation of local Lorentz invariance \cite{Lorentz4}. To be
consistent with those experiments, the violation of the local
Lorentz invariance in $f(T)$ gravity should be very small. Note that
$f''(T)$ can be used as a parameter to denote the violation of the
local Lorentz invariance since it vanishes when $f(T)$ gravity
reduces to $TG$ with local Lorentz invariance. So $f''(T)$ is also
expected to be very small and in that case the entropy production
term eq.~(\ref{entropy pro}) can be ignored. Thus, for a small
$f''(T)$, the first law of black holes is satisfied approximatively
and the entropy is $\frac{f'(T)}{4}A$.

Finally, we observe that for the special case $f'(T)=1$ the entropy
production vanishes and the entropy reduces to that of Einstein
gravity $S=\frac{A}{4}$, which is consistent with the equivalence
between $TG$ and Einstein gravity.

\section{Reason for violation of First law of $f(T)$ gravity}

In this section, we search for the reason for violation of first law
of black holes in $f(T)$ gravity by investigating two concrete
examples, Rindler space and Minkowski space. We find that it is the
violation of the local Lorentz invariance that leads to the
breakdown of first law of black holes in $f(T)$ gravity. Although
all the matter fields with the local Lorentz invariance see the same
horizon and Hawking temperature, some gravitational degrees of
freedom in $f(T)$ gravity feel a different background metric,
horizon and Hawking temperature. Black holes in such a situation
cannot be in equilibrium \cite{Lorentzviolation,Dubovsky,Jacobson2}
and consequently the first law in equilibrium is violated.

For simplicity, we focus on $f(T)$ gravity in $3D$ below (The
discussions below can be easily extended to the $4D$ case.). As the
first example, let us consider the linear perturbation equations of
$f(T)$ gravity with the background tetrad fields $^{0}e_{a\mu}={\rm
diag}(x,1,1)$ and perturbations $^{1}e_{a\mu}$. The background
spacetime is Rindler space with metric $ds^2=-x^2dt^2+dx^2+dy^2$,
while the perturbations of metric are
$h_{\mu\nu}={}^{0}e_{a\mu}\eta^{ab}\
^{1}e_{b\nu}+{}^{0}e_{a\nu}\eta^{ab}\ ^{1}e_{b\mu}$. For the sake of
convenience, we set $f(0)=0$, which means that there is no
cosmological constant term in the action of $f(T)$ gravity. Note
that the background tetrad fields $^{0}e_{a\mu}$ satisfy field
equations of $f(T)$ gravity in vacuum, and the background torsion
scalar $^0T=0$.

We recall that the equation of motion of $f(T)$ gravity is
eq.~(\ref{eom}), from which the linear perturbation equation  can be
derived in terms of $^0T=0$,
\begin{eqnarray}\label{linear}
\frac{f'(0)}{2}[\nabla_{\nu}\nabla^\rho
\bar{h}_{\rho\mu}+\nabla_{\mu}\nabla^\rho \bar{h}_{\rho\nu}-\Box
\bar{h}_{\mu\nu}-\
^0g_{\mu\nu}\nabla^\rho\nabla^\sigma\bar{h}_{\rho\sigma}]+f''(0)\,^0S_{
\nu\mu}^{\ \ \rho}\,\nabla_\rho\, ^1T=8\pi\,{}^1\Theta_{\mu\nu},
\end{eqnarray}
where $\nabla_\mu$ is the covariant derivative defined by
$^0g_{\mu\nu}$, and $\Box=\nabla_\mu\nabla^\mu$. $^1T$ and
$^1\Theta_{\mu\nu}$ are the perturbations of torsion scalar and
stress tensor, respectively.
$\bar{h}_{\mu\nu}=h_{\mu\nu}-\frac{h}{2}\ ^0g_{\mu\nu}$ and
$h=h_{\mu\nu}\ ^0g^{\mu\nu}$. Similar to Einstein gravity, we can
impose the Lorentz gauge $\nabla^\mu \bar{h}_{\mu\nu=0}$ to simplify
the above equation. The reason is that $f(T)$ gravity is also
invariant under the general coordinate transformations,
$x^\mu\rightarrow x^\mu+\zeta^\mu$, $h_{\mu\nu}\rightarrow
h_{\mu\nu}+2\nabla_{(\mu}\zeta_{\nu)}$. For every given
$h_{\mu\nu}$, we can always find some suitable gauge parameters
$\zeta^\mu$ to make
$h'_{\mu\nu}=h_{\mu\nu}+2\nabla_{(\mu}\zeta_{\nu)}$ satisfy the
Lorentz gauge $\nabla^\mu \bar{h'}_{\mu\nu=0}$. In fact, we only
need to solve the equation for $\zeta_{\mu}$,
$-\nabla^\mu\bar{h}_{\mu\nu}=\Box\zeta_{\nu}+R_{\nu\mu}\zeta^{\mu}=\Box\zeta_{\nu}$,
where we have used $R_{\mu\nu}=0$ in Rindler space. It is clear that
solutions always exist for the above equation. Applying the Lorentz
gauge $\nabla^\mu \bar{h}_{\mu\nu}=0$, we can simplify
eq.~(\ref{linear}) as follows:
\begin{eqnarray}\label{linear1}
-\frac{f'(0)}{2}\Box \bar{h}_{\mu\nu}+f''(0)\,^{0}S_{\nu\mu}^{\ \
\rho}\,\nabla_\rho\, ^{1}T=8\pi\, ^{1}\Theta_{\mu\nu}.
\end{eqnarray}
Note that $f''(0)$ can be used to denote the violation of local
Lorentz invariance, and that when it vanishes the above perturbation
equation recovers the local Lorentz invariance. Using background
tetrad fields $^{0}e_{a\mu}={\rm diag}(x,1,1)$, we can derive
$^{0}S_{\nu\mu}^{\ \ \rho}$ (see eq.~(\ref{dualtorsion})). The
non-zero results are given by
\begin{eqnarray}
{}^{0}S_{yx}^{\ \ y}&=&-\frac{1}{x}\label{S1},\\
{}^{0}S_{yy}^{\ \ x}&=&\frac{1}{x}\label{S2}.
\end{eqnarray}
From the antisymmetric part of eq.~(\ref{linear1}), $S_{[\nu\mu]}^{\
\ \ \ \rho}\,\nabla_\rho\, ^{1}T=0$, and eq.~(\ref{S1}), we can
derive
\begin{eqnarray}
\partial_y\, ^1T=0.
\end{eqnarray}
As the simplest solution of the above equation, we require the
perturbation $^1e_{a\mu}$ be independent of coordinate $y$.
Substituting eq.~(\ref{S2}) into eq.~(\ref{linear1}), we find that
most components of $\bar{h}_{\mu\nu}$ obey the same equation as that
in Einstein gravity,
\begin{eqnarray}\label{Rindler1}
-\frac{f'(0)}{2}\Box \bar{h}_{\mu\nu}=8\pi\, ^{1}\Theta_{\mu\nu},
\qquad f'(0)=1,
\end{eqnarray}
except for $\bar{h}_{yy}$. For those fields that satisfy the same
equation as that in Einstein gravity, they feel the same background
metric (Rindler space in our case), therefore see the same horizon
and Hawking temperature as the matter fields.

However, $\bar{h}_{yy}$ satisfies a different equation in the form
of
\begin{eqnarray}\label{Rindler2}
-\frac{f'(0)}{2}\Box \phi+f''(0)\frac{1}{x}\partial_x\, ^1T=8\pi\,
^{1}\Theta',
\end{eqnarray}
where $\phi$ stands for $\bar{h}_{yy}$, and
${}^1\Theta'={}^1\Theta_{yy}$. Note that $\bar{h}_{yy}$ behaves like
a scalar under the action of $\Box$ in Rindler space, $\Box
\bar{h}_{yy}=\frac{1}{\sqrt{-g}}\partial_\nu(\sqrt{-g}g^{\nu\mu}\partial_{\mu}\bar{h}_{yy})$,
thus we denote it by $\phi$. For simplicity, we require that all
$^1e_{a\mu}$ vanish except for $^1e_{(2)y}=\phi$. Consequently, we
have $h_{yy}=2\phi$, $\bar{h}_{yy}=\phi$ and $^1T=2\
^0S^{a\mu\nu}\partial_\mu\, ^1e_{a\nu}-2\ ^0S^{avc}\ ^0T_{ adc}\
^1e^{d}_{\ \nu}=-\frac{2}{x}\partial_x \phi$. In view of
$\partial_y\, ^1e_{a\mu}=0$, we observe that this choice satisfies
the Lorentz gauge $\nabla^\mu\bar{h}_{\mu\nu}$. Now,
eq.~(\ref{Rindler2}) becomes
\begin{eqnarray}\label{Rindler3}
-\frac{f'(0)}{2}\Box
\phi-2f''(0)\left(\frac{1}{x}\partial_x\right)^2\phi=8\pi\,
^{1}\Theta'.
\end{eqnarray}
It should be stressed that for our simple choice that all
$^1e_{a\mu}$ vanish except for $^1e_{(2)y}=\phi$, we have
$\bar{h}_{tt}=x^2\phi$ and $\bar{h}_{xx}=-\phi$, which leads to two
constraints for $\Theta_{tt}$ and $\Theta_{xx}$ from
eq.~(\ref{Rindler1}). For simplicity, we require that $\Theta_{tt}$
and $\Theta_{xx}$ satisfy the constraints.

Redefine $\phi=\sqrt[4]{\frac{x^2}{\mid \epsilon+x^2\mid}}\
\bar{\phi}$ and $\Theta'=\sqrt[4]{\frac{x^2}{\mid
\epsilon+x^2\mid}}\ \bar{\Theta}'$, where
$\epsilon=\frac{4f''(0)}{f'(0)}$, we can rewrite the above equation
as
\begin{eqnarray}\label{Rindler3}
-\frac{f'(0)}{2}[\bar{\Box} - V(x)]\bar{\phi}=8\pi\,
^{1}\bar{\Theta}',
\end{eqnarray}
where $V(x)=\frac{\epsilon(3\epsilon+4x^2)}{4x^4(\epsilon+x^2)}$,
and $\bar{\Box}$ is defined by the effective metric
$\bar{g}_{\mu\nu}$  which takes the form
\begin{equation}\label{g}
\bar{g}_{\mu\nu}=\left(
\begin{array}{ccc}
 -x^2 & 0 & 0 \\
  0& \frac{x^2}{x^2+\epsilon} & 0\\
 0 & 0 & 1
\end{array}
\right).
\end{equation}
As a result, the field $\bar{\phi}$ feels an effective metric
$\bar{g}_{\mu\nu}$ different from that of Rindler space.

If $\epsilon>0$, the horizon of this effective metric still lies at
$x=0$, and the Hawking temperature is
\begin{eqnarray}\label{T1}
T_1=\frac{1}{2\pi}N_{\mu}\nabla^{\mu}e^\varphi=\frac{1}{2\pi
x}\sqrt{\epsilon+x^2},
\end{eqnarray}
where $N_\mu=(0,\frac{x}{\sqrt{ \epsilon+x^2}},0)$ is a unit outward
pointing vector normal to the horizon,
$\varphi=\frac{1}{2}\log(-\zeta^\mu\zeta_\mu)$ is the Newton's
potential and $\zeta^\mu=(1,0,0)$ is a time-like Killing vector.
Note that the temperature $T_1$ diverges at the horizon, and the
worse is that there is a naked singularity at $x=0$ in view of Ricci
scalar $\bar{R}=\frac{2\epsilon}{x^4}$. According to the cosmic
censorship conjecture, no naked singularities other than the Big
Bang singularity exist in the universe. Therefore, in order to avoid
the naked singularities and divergence of temperature, $\epsilon$
would not be positive.

For $\epsilon<0$, the position of the horizon turns to be
$x=\sqrt{-\epsilon}$, where $\bar{R}=\frac{2\epsilon}{x^4}$,
$\bar{R}^{\mu\nu\rho\sigma}\bar{R}_{\mu\nu\rho\sigma}=\frac{4\epsilon^2}{x^8}$
and $\bar{R}^{\mu\nu}\bar{R}_{\mu\nu}=\frac{2\epsilon^2}{x^8}$ have
a good behavior and the singularity at $x=0$ is hidden within the
horizon. The temperature $T_2=0$ on the horizon $x=\sqrt{-\epsilon}$
can be read out from eq.~(\ref{T1}), which is different from the
temperature $T=\frac{1}{2\pi}$ felt by matter fields in Rindler
space.

Now let us summarize our results. At first, the scalar field
$\bar{\phi}$ in $f(T)$ gravity feels an effective metric
eq.~(\ref{g}) different from that felt by matter fields, it
therefore sees a different horizon and Hawking temperature. Black
holes in such a situation would not be in the equilibrium state.
Second, notice that the parameter $\epsilon=\frac{4f''(0)}{f'(0)}$
is related to the violation of local Lorentz invariance.
 When $\epsilon$ vanishes, eq.~(\ref{linear}) recovers
the local Lorentz invariance and the effective metric eq.~(\ref{g})
reduces to the metric of Rindler space. Furthermore, when $f''(T)=0$
and $\epsilon=0$, the entropy production terms (eq.~(\ref{entropy
pro})) vanish and the first law of black hole thermodynamics
recovers. As a result, the breakdown of first law of black holes
results from the violation of local Lorentz invariance
($\epsilon\neq 0$). At last, $\epsilon$ should be negative in order
to avoid the naked singularity.

To end up this section, we briefly discuss the second example with
background metric $^0g_{\mu\nu}={\rm diag}(-1,1,1)$ and tetrad
fields
\begin{equation}\label{tetead}
^0e_{a\mu}=\left(
\begin{array}{ccc}
 \cosh(x) & \sinh(x) & 0 \\
 \sinh(x) & \cosh(x) & 0 \\
 0 & 0 & 1
\end{array}
\right).
\end{equation}
Again, $^0e_{a\mu}$ satisfy the field equations of $f(T)$ gravity in
vacuum when $f(0)=0$. Note that $^0T=0$ and the non-vanishing
${}^0S_{\mu\nu}^{\ \ \, \rho}$ are $^{0}S_{yt}^{\ \
y}={}^{0}S_{yy}^{\ \ t}=1$. After imposing the Lorentz gauge
$\partial_\mu \bar{h}^\mu_{\ \nu}=0$, we conclude that most of
metric perturbations $\bar{h}_{\mu\nu}$ obey the same equation
eq.~(\ref{Rindler1}) as that in Einstein gravity expect for
$\bar{h}_{yy}$ which satisfies
\begin{eqnarray}\label{flatspace}
-\frac{f'(0)}{2}\Box \phi+f''(0)\partial_t \, ^{1}T=8\pi\,
^{1}\Theta_{yy},
\end{eqnarray}
where $\phi$ denotes $\bar{h}_{yy}$. Focusing  on the case all
perturbations of tetrad fields vanish expect for $^1e_{(2)y}=\phi$,
we have $^1T=-2\partial_t\phi$. Thus, the above equation becomes
\begin{eqnarray}\label{flatspace1}
-\frac{f'(0)}{2}[\Box \phi+\epsilon(\partial_t)^2 \phi ]=8\pi\,
^{1}\Theta_{yy},
\end{eqnarray}
from which one can easily read out the effective metric
$\bar{g}_{\mu\nu}$ as follows:
\begin{equation}\label{g1}
\bar{g}_{\mu\nu}=\left(
\begin{array}{ccc}
 \frac{-1}{1-\epsilon}& 0 & 0 \\
 0& 1 & 0\\
 0 & 0 & 1
\end{array}
\right).
\end{equation}
From $ds^2=\bar{g}_{\mu\nu}dx^\mu dx^\nu=0$, we get the speed of
field $\phi$, $v=\frac{1}{\sqrt{1-\epsilon}}$. It is interesting
that if we require that $v$ does not exceed the speed of light, we
get $\epsilon<0$, which is the same as the condition in the first
example given for getting rid of the naked singularity. It should be
mentioned that one can derive a similar condition in light of the
recent work of Y. F. Cai et al. \cite{Cai}. From eq.~(28) of their
paper \cite{Cai}, we note that both $f''(T)$ and
$\epsilon=\frac{4f''(0)}{f'(0)}$ ($f'(0)>0$ from
eq.~(\ref{constraint})) should be negative if we require the sound
speed parameter $c_s$ does not exceed the speed of light.

\section{Conclusion}

In this paper, we have shown that, in general, the first law of
black hole thermodynamics $\delta Q=T\delta S$ is violated in $f(T)$
gravity, and only for some special cases can it be recovered. There
is entropy production even in the static spacetime, and there are
strict constraints for $f(T)$ gravity in order to maintain the
positivity of entropy and entropy production. We find that the
violation of first law results from the lack of local Lorentz
invariance in $f(T)$ gravity. Through investigating two concrete
examples, we observe that the effective metric felt by some degrees
of freedom in $f(T)$ gravity is different from the background metric
felt by matter fields because of the violation of local Lorentz
invariance. The degrees of freedom therefore see a different horizon
and Hawking temperature. Black holes in such a situation would not
be in equilibrium, so it is the violation of local Lorentz
invariance that leads to the breakdown of the first law of black
hole thermodynamics, $\delta Q=T\delta S$, in $f(T)$ gravity. To
avoid the naked singularity and super velocity of light in the two
examples, we get the condition $\epsilon=\frac{4f''(0)}{f'(0)}<0$,
where $\epsilon$ is a parameter which denotes the violation of local
Lorentz invariance in $f(T)$ gravity. To be consistent with
experiments, $\epsilon$ and $f''(T)$ should be small. In that case,
the entropy production term is small compared with the first term in
the last line of eq.~(\ref{Q5}), thus the first law of black hole
thermodynamics can be satisfied approximatively and the entropy of
black holes in $f(T)$ gravity equals $\frac{f'(T)}{4}A$
approximatively.

\section*{Acknowledgements}
R-X.M. would like to thank T. Wang for useful discussions. M.L. and
R-X.M. are supported by the NSFC grants No.10535060, No.10975172 and
No.10821504, and by the 973 program grant No.2007CB815401 of the
Ministry of Science and Technology  of China. Y-G.M. is supported by
the NSFC grant No.11175090.

\section*{Appendix A}
We give the proof that the entropy production on the null surface is
indeed non-vanishing by studying a specific example, Rindler space.
The metric of Rindler space is
\begin{eqnarray}\label{Rindler}
ds^2=-x^2dt^2+dx^2+dy^2+dz^2,
\end{eqnarray}
where $x\in[0, \infty)$. We choose the following tetrad fields
$e_{a\mu}$,
\begin{equation}\label{background1}
\left(
\begin{array}{cccc}
x& 0& 0& 0 \\
0& \cos{[g(x)y]}& \sin{[g(x)y]}& 0 \\
0& -\sin{[g(x)y]}& \cos{[g(x)y]}& 0 \\
0& 0& 0& 1 \\
\end{array}
\right)
\end{equation}
with an arbitrary function $g(x)$ and the torsion scalar
$T=-\frac{2g(x)}{x}$. One can check that they satisfy the equations
of motion eqs.~(\ref{motion1}) and (\ref{motion10}) as long as the
matter stress-tensor is given by
\begin{eqnarray}\label{stress}
\Theta^{0}_0&=&\frac{1}{16\pi x^2}\left[x^2f(T)+2x g f'(T)+4g(g-x
g')f''(T) \right],\nonumber\\
\Theta^1_1&=&\frac{1}{8\pi}\left[\frac{f(T)}{2}+\frac{g
f'(T)}{x}\right],\nonumber\\
\Theta^2_2&=&\frac{1}{8\pi}\left[\frac{f}{2}+\frac{x^2 g f'(T)+2(g-x
g')f''(T)}{x^3}\right],\nonumber\\
\Theta^3_3&=&\frac{1}{8\pi}\left[\frac{f}{2}+\frac{g
f'(T)}{x}+\frac{2(1+x g)(g-x g')f''(T)}{x^3}\right].
\end{eqnarray}
One can always choose suitable functions $g(x)$ and $f(T)$ to make
$\Theta^\mu_\nu$ be regular and simultaneously to keep the entropy
production non-vanishing. For example, set
$g(x)=\frac{1}{2}x^{3}e^{-|x|}$ and $f(T)=\sum\limits_{n=1}^{N} a_n
T^n$ ($T=-x^2e^{-|x|}$), where $N$ is an arbitrary finite integer
greater than 1, we find that the matter stress-tensor
eq.~(\ref{stress}) is regular in the whole space.

It should be stressed that there are many different choices of null
surface in Rindler space. Without the loss of generality, we focus
on the null surface $x=e^{t}$ below. The corresponding Killing
vector and null vector on this null surface are
$\xi^{\mu}=(1-\frac{\cosh t}{x},\sinh t,0,0)$ and
$k^{\mu}\sim(\frac{e^t}{x^2},1,0,0)$, respectively. One can check
that $\xi^\mu\xi_\mu=\xi^\mu k_\mu=k^\mu k_\mu=0$, $\xi^\mu\sim
k^\mu$ on this null surface. From eq.~(\ref{entropy pro}), we find
that the entropy production on the null surface $x=e^{t}$ is
proportional to
\begin{eqnarray}\label{production1}
-k^{\nu}\nabla^\mu f'(T)(\xi^\rho S_{\rho \nu\mu}-\nabla_\nu
\xi_\mu)\sim -2f''(T)\frac{e^{t}(x g'-g)[(x- \cosh t)g+1]}{x^3},
\end{eqnarray}
which is non-vanishing generally.

\section*{Appendix B}

We give an exact solution of eqs.~(\ref{motion1}) and
(\ref{motion10}) which has the properties of the usual Rindler
space. All the modes feel the same Rindler space metric as that felt
by matter fields and no modes can escape from inside of the horizon.
Similar to Sect. IV, in order to get the effective metric felt by
the tetrad fields, let us investigate the linear perturbation
equations of $f(T)$ gravity with the background tetrad fields
\begin{equation}\label{background2}
{}^{0}e_{a\mu}=\left(
\begin{array}{cccc}
x \cosh t& \sinh t& 0& 0 \\
x \sinh t& \cosh t& 0& 0 \\
0& 0& 1& 0 \\
0& 0& 0& 1 \\
\end{array}
\right)
\end{equation}
and perturbations $^{1}e_{a\mu}$. The background spacetime is
Rindler space with metric $ds^2=-x^2dt^2+dx^2+dy^2+dz^2$, while the
perturbations of metric are $h_{\mu\nu}={}^{0}e_{a\mu}\eta^{ab}\
^{1}e_{b\nu}+{}^{0}e_{a\nu}\eta^{ab}\ ^{1}e_{b\mu}$. Notice that we
have
\begin{equation}\label{eq1}
{}^0T=0,\qquad {}^0S_{\mu\nu}^{\ \ \ \rho}=0,
\end{equation}
for the background tetrad fields eq.~(\ref{background2}).
Eq.~(\ref{background2}) satisfies eqs.~(\ref{motion1}) and
(\ref{motion10}) provided the background matter stress-tensor is
${}^0\Theta^{\mu}_{\nu}=\frac{f(0)}{16\pi}\delta^{\mu}_{\nu}$. Using
eq.~(\ref{eq1}), we can easily find that the linear perturbation of
eq.~(\ref{motion10}) automatically vanishes. Thus, we only need to
study the linear perturbation of eq.~(\ref{motion1}). Following the
approach of Sect. IV, we can derive
\begin{eqnarray}\label{eq2}
-\frac{f'(0)}{2}\Box \bar{h}_{\mu\nu}=8\pi\, ^{1}\Theta_{\mu\nu},
\end{eqnarray}
which is exactly the same as the linear perturbation equations of
Einstein gravity if we set $f'(0)=1$. Let us recall that
$\bar{h}_{\mu\nu}=h_{\mu\nu}-\frac{h}{2}\ {}^0g_{\mu\nu}$ and
${}^{1}\Theta_{\mu\nu}$ is the linear perturbation of the matter
stress-tensor. From eq.~(\ref{eq2}), it is clear that all the modes
of tetrad fields feel the same metric as the background spacetime,
Rindler space. Thus, at least in the linear perturbation, all the
modes feel the same horizon and null surfaces, and no modes can
escape from inside of the horizon. From eqs.~(\ref{Q5}) and
(\ref{eq1}), it is interesting to note that the first law $\delta
Q=T\delta S$ recovers on the null surface of Rindler space with
tetrad fields eq.~(\ref{background2}), and the entropy on the null
surface can be read out from eq.~(\ref{Q5}) as $S=\frac{f'(0)}{4}A$.

\end{document}